%% file: IRS_DFRC_tutorial.tex
\documentclass[conference]{IEEEtran}

\IEEEoverridecommandlockouts
\usepackage{cite}
\usepackage{amsmath,amssymb,amsfonts,comment}
\usepackage{algorithmic}
\usepackage{graphicx}
\usepackage{textcomp}
\usepackage{xcolor,balance}
\usepackage[ruled,vlined]{algorithm2e}
\def\BibTeX{{\rm B\kern-.05em{\sc i\kern-.025em b}\kern-.08em
  T\kern-.1667em\lower.7ex\hbox{E}\kern-.125emX}}

\newcommand{\tr}[1]{\mathrm{tr}\!\left[{#1}\right]}

\begin{document}

\IEEEoverridecommandlockouts
\IEEEpubid{\begin{minipage}{\textwidth}\ \\[10pt]
        \normalsize{979-8-3503-3959-8/23/\$31.00 \copyright 2023 IEEE}
\end{minipage}}

\title{Efficient Beamforming Designs for IRS-Aided DFRC Systems   
}

\author{
	\IEEEauthorblockN{Yi-Kai Li and Athina Petropulu}
	\IEEEauthorblockA{Dept. of Electrical and Computer Engineering, Rutgers University, Piscataway, NJ, USA
		\\
		E-mail: \{yikai.li, athinap\}@rutgers.edu\vspace{-0mm}}
		\thanks{This work was supported by ARO grants W911NF2110071 and W911NF2320103.}
		}

\maketitle

\begin{abstract}
This short tutorial presents several ideas for designing dual function radar communication (DFRC) systems aided by intelligent reflecting surfaces (IRS). These   problems are highly nonlinear in the   IRS parameter matrix, and further, the IRS parameters are subject to non-convex unit modulus constraints. We present classical semidefinite relaxation based methods, low-complexity minorization based optimization methods, low-complexity Riemannian manifold optimization methods, and near optimal branch and bound based methods.

\end{abstract}


\section{INTRODUCTION}
The wireless spectrum becomes increasingly congested and spectrum reuse is necessary to achieve high spectral efficiency. How to share the spectrum between radar and communication systems, has attracted significant attention for decades \cite{Li2016optimum}. Recently, dual function radar communication (DFRC) systems \cite{Zhang2021anoverview,Liu2020joint,Xu2023abandwidth} provide a new paradigm for spectrum sharing.
DFRC systems reuse the transmit waveform and share a common hardware platform between the radar and communication functionalities. As a result, DFRC systems achieve high spectral efficiency and reduced hardware cost.

In order to have access to higher bandwidth, which is needed for higher sensing resolution and higher communication rates, 
future wireless networks will need to use high frequencies.  However, such frequencies experience high attenuation.
 Intelligent reflecting surface (IRS)
 have the potential to create a smart propagation environment, 
 and combat the effects of attenuation.
 An IRS
 is a passive array composed of elements of printed dipoles, each of which can change the phase of the impinging electromagnetic wave in a computer controlled fashion \cite{Wu2019intelligent,Wu2020towards}. These elements can cooperatively beamform towards  desired directions, or avoid  undesired directions. When used with a DFRC system, an IRS platform can create extra paths between the DFRC radar the communication users and the radar targets. Thereby, IRS can assist the sensing and communication functions of the DFRC system simultaneously, by customizing the wireless propagation environment for the radar targets \cite{Buzzi2022foundations} and communication users \cite{Wu2019intelligent}.

The design of  IRS-assisted 
DFRC  systems is a challenging non-convex problem. The radar precoder and the IRS parameters are coupled variables, and need to be jointly designed. In addition,  the design problem consists of non-convex objectives, non-convex manifold constraints, and high order functions in the objective or constraints \cite{Li2023minorization,Li2022_sam,Jiang2021intelligent,Liu2022joint}.

In this paper we present different methods for addressing the challenging  design of IRS-aided DFRC systems. We start with classical semidefinite relaxation (SDR) based optimization methods. Then, we present two closed-form expression based   algorithms, which are applicable to   large IRS scenarios, where the design complexity is high. One of them is based on minorization or MM \cite{Sun2017majorization}, and the other one is based on Riemannian manifold optimization \cite{Li2022_sam}. We should note that large size IRS are needed in order to  
achieve high beamforming gain \cite{Najafi2021physics}.  Finally,  we present benchmark near-optimal branch-and-bound based algorithm to enhance the system performance by iteratively selecting the current best solution from local optima.

\vspace{3mm}
\section{SYSTEM MODEL}
\label{sec:models}

\begin{figure}[!t]\vspace{0mm} 
	\hspace{12mm} 
	\def\svgwidth{180pt} 
	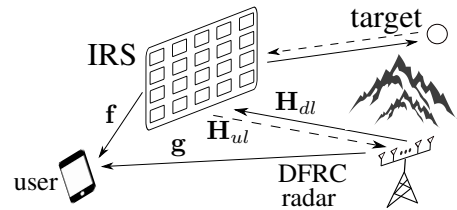 
	\caption{IRS-assisted DFRC system.}\vspace{-5mm} 
	\label{fig:system_model}
\end{figure}

Let us consider the IRS-assisted DFRC system  shown Fig. \ref{fig:system_model}. The collocated DFRC transmitter and receiver are respectively modeled as $N_T$- and $N_R$-element uniform linear arrays (ULAs). For both the inter-element space is $d$. The radar transmits a single waveform, which  is designed to track a non-line-of-sight (NLOS) point target,  while simultaneously conveying information to a single-antenna communication user. The system is assisted by an   $N$-element IRS. Perfectly known flat fading channels are assumed. The signal emitted from the DFRC radar is  

\vspace{-5mm}
\begin{eqnarray}
	\mathbf x = \mathbf w  s,
\end{eqnarray}
\vspace{-5mm}

\noindent where $\mathbf w \in \mathbb C^{N_T \times 1}$ is the radar precoding vector, and $s$ is the radar probing  waveform that also contains communication information. $s$ is assumed to be zero-mean while with unit variance.
%
Since there is no direct path to the target, the transmitted signal reaches the target after being reflected by the IRS, then it is reflected by the target, and arrives at the radar receiver after being reflected by the IRS again. The signal at the radar receiver can be expressed as

\vspace{-6mm}
\begin{small}
\begin{eqnarray} \label{eqn:y_R}
   \mathbf y_R &=& 
    {\beta \mathbf H_{ul} \mathbf \Phi \mathbf a_I(\psi_{a},\psi_{e}) \mathbf a_I^{T}(\psi_{a},\psi_{e}) \mathbf \Phi \mathbf  H_{dl} }
    \mathbf x+ \mathbf n_R \nonumber\\  &=& \mathbf C_{T} \mathbf x + \mathbf n_R, 
\end{eqnarray}
\end{small}
\vspace{-7mm}

\noindent where $\beta$ is the   channel corresponding to  the radar-IRS-target-IRS-radar path; $\mathbf H_{ul}$ and $\mathbf H_{dl}$  are respectively the normalized  IRS-radar and  radar-IRS channels; $\mathbf \Phi = \text{diag}([e^{j \phi_1},\cdots,e^{j \phi_n},\cdots,e^{j \phi_N}])$  is a diagonal matrix denoting the IRS parameter, with $\phi_n \in [0,2 \pi)$ denoting the phase shift induced by the $n$-th IRS element; $\mathbf a_I(\psi_a,\psi_e)$ is the IRS steering vector, where $\psi_a$ and $\psi_e$ are respectively the target angles relative to the IRS in the azimuth and elevation planes; $\mathbf n_R \sim \mathcal{CN}(\mathbf 0_{N_R \times 1}, \sigma_R^2 \mathbf I_{N_R})$ is the additive white Gaussian noise (AWGN) at the radar receiver, where $\sigma_R^2$ is the  noise power at each radar receive antenna.

The received signal at the communication user is

\vspace{-6mm}
\begin{small}
\begin{eqnarray} \label{eqn:y_U}
	 y_{U} = (\beta_H^{\frac{1}{2}} \mathbf f^T \mathbf \Phi \mathbf H_{dl} + \mathbf g^T) \mathbf w s +  n_U = \mathbf c_U^T \mathbf w s +  n_U,
\end{eqnarray}
\end{small}
\vspace{-5mm}

\noindent where $\beta_H$ is the path-loss of the radar-IRS channel,  $\mathbf f \in \mathbb C^{N \times 1}$ is the   user-IRS channel, $\mathbf g \in \mathbb C^{N_T \times 1}$ is the user-radar channel, and $n_U \sim \mathcal {CN}(0,\sigma_U^2)$ is   AWGN  at the user.

The SNR at radar receiver and communication user are

\vspace{-6mm}
\begin{small}
\begin{eqnarray}	&&\!\!\!\!\!\!\!\!\!\!\!\!\!\!\!\gamma_R =  (1/{\sigma_R^2}) \tr{\mathbf C_T \mathbf w \mathbf w^H \mathbf C_T^H}, \label{eqn:gamma_R}\\&&\!\!\!\!\!\!\!\!\!\!\!\!\!\!\!
	\gamma_U =  (1/{\sigma_U^2}) |\mathbf c_U^T \mathbf w|^2. \label{eqn:gamma_U}
\end{eqnarray}
\end{small}

\section{SYSTEM DESIGN}
\label{sec:pagestyle}
Let us design the radar precoder $\mathbf w$ and IRS parameter $\mathbf \Phi$ so that we maximize the  weighted sum of the SNRs at  the radar receiver and user,  while meeting certain constraints, i.e., 

\vspace{-6mm}
\begin{small}
\begin{subequations} \label{eqn:orig_prob}
	\begin{eqnarray} 
		\max_{\mathbf w, \mathbf \Phi}&&	\;\;\;\;	\alpha\gamma_R + (1-\alpha) \gamma_U\label{eqn:obj}\\ \mathrm{s.t.} && \;\;\;\; \tr{\mathbf w \mathbf w^H} = P_R \label{eqn:tot_power}\\&& \;\;\;\; \|\mathbf w \mathbf w^H-\mathbf R_D\|_F^2 \leq \gamma_{BP} \label{eqn:beam_pattern_cons}\\&& \;\;\;\; 
		|\mathbf \Phi_{n,n}| = 1,\;\; \forall n = 1, \cdots, N\label{eqn:unit_modu}
	\end{eqnarray}
\end{subequations}
\end{small}
\vspace{-6mm}

\noindent where the real constant $\alpha$ is a weight factor;  (\ref{eqn:tot_power}) ensures that the power of the transmitted waveform stays within the power budget,  $P_R$; (\ref{eqn:beam_pattern_cons}) ensures  that the maximum beampattern deviation from a desired one is less than a threshold $\gamma_{BP}$, with $\mathbf R_D$ denoting  the covariance matrix of the transmitted waveform; (\ref{eqn:unit_modu}) implies that each element of $\mathbf \Phi$ has unit modulus. The latter constraint models the fact   that the IRS does not have any radio frequency (RF) chains and thus cannot change the magnitude of the reflected signal. Therefore,  it only changes the phase of the impinging signal.


Problem (\ref{eqn:orig_prob}) is highly non-convex due to the following facts. (i) The variables $\mathbf w$ and $\mathbf \Phi$ are coupled with each other. (ii) The radar SNR, $\gamma_R$, is fourth order in $\mathbf \Phi$, since $\gamma_R$ is quadratic into $\mathbf C_T$ per Eq. (\ref{eqn:gamma_R}), where   $\mathbf C_T$ is quadratic in $\mathbf \Phi$ per Eq. (\ref{eqn:y_R}). (iii)  (\ref{eqn:unit_modu}) is non-convex unit modulus constraint (UMC) on $\mathbf \Phi$.

To decouple $\mathbf w$ and $\mathbf \Phi$, we can separate (\ref{eqn:orig_prob}) into to two sub-problems, i.e., optimize with respect to $\mathbf w$ by fixing $\mathbf \Phi$, and optimize with respect to $\mathbf \Phi$ by fixing $\mathbf w$. The two sub-problems are solved alternatingly till convergence is reached \cite{Li2017_coexistence}.

\subsection{First sub-problem: Solve for $\mathbf w$ with fixed $\mathbf \Phi$}

The first sub-problem can be casted as follows:

\vspace{-5mm}
\begin{small}
\begin{subequations} \label{eqn:problem1}
	\begin{eqnarray} 
	\!\!\!\!\!\!\!\!\!(\mathrm P_{w})	\qquad \max_{\mathbf w }&&	\;\;\;\;	\alpha\gamma_R + (1-\alpha) \gamma_U\label{eqn:obj}\\ \!\!\!\!\!\!\!\!\!\mathrm{s.t.} && \;\;\;\; \tr{\mathbf w \mathbf w^H} = P_R \label{eqn:tot_power1}\\\!\!\!\!\!\!\!\!\!&& \;\;\;\; \|\mathbf w \mathbf w^H-\mathbf R_D\|_F^2 \leq \gamma_{BP} \label{eqn:beam_pattern_cons1}
	\end{eqnarray}
\end{subequations}
\end{small}
\vspace{-5mm}

\noindent Both $\gamma_R$ and $\gamma_U$ are quadratic functions of $\mathbf w$, per Eq. (\ref{eqn:gamma_R}) and (\ref{eqn:gamma_U}). Therefore, the 
 objective  of weighted sum of $\gamma_R$ and $\gamma_U$, in Eq. (\ref{eqn:obj}), is quadratic into $\mathbf w$. Problem (\ref{eqn:problem1}) is a quadratic programming problem for variable $\mathbf w$, which can be solved by 
 SDR \cite{Luo2010semidefinite}, or majorization minimization (MM) \cite{Sun2017majorization}. 
 
 In SDR method \cite{Luo2010semidefinite}, an auxiliary variable, $\mathbf R_w =  \mathbf w^H \mathbf w$, is defined. The objective and constraints are thereby linear functions of $\mathbf R_w$, which forms a linear programming problem in variable $\mathbf R_w$. The optimal solution, say $\mathbf R_w^{\ast}$, can be readily obtained via numerical solvers like CVX \cite{cvx}. $\mathbf w$ can be recovered from $\mathbf R_w^{\ast}$ by Gaussian randomization \cite{Luo2010semidefinite}, i.e., multiple samples of random vectors with  covariance  $\mathbf R_w^{\ast}$, $\mathbf \xi \sim \mathcal{CN}(\mathbf 0, \mathbf R_w^{\ast})$ are generated and the  one that maximizes the objective is chosen as the optimal  $\mathbf w$. In the MM method \cite{Sun2017majorization}, linear surrogate functions can be created for the quadratic functions in the objective and constraints in (\ref{eqn:problem1}), which results in a linear programming problem.
 
\subsection{Second sub-problem: Solve for $\mathbf \Phi$ with fixed $\mathbf w$}\label{sec:problem2}
The second sub-problem is given as

\vspace{-5mm}
\begin{small}
\begin{subequations} \label{eqn:problem2}
	\begin{eqnarray} 
	(\mathrm P_{\Phi}) \qquad	\max_{\mathbf \Phi}&&	\;\;\;\;	\alpha\gamma_R + (1-\alpha) \gamma_U\label{eqn:obj2}\\ \mathrm{s.t.} && \;\;\;\; 
		|\mathbf \Phi_{n,n}| = 1,\;\; \forall n = 1, \cdots, N\label{eqn:unit_modu2}
	\end{eqnarray}
\end{subequations}
\end{small}

The difficulty and challenge of designing the IRS DFRC system mainly lies in the optimization of the objective function with respect to IRS parameters subject to the highly non-convex UMCs. 
Note that $\gamma_U$ in (\ref{eqn:obj2}) is quadratic into $\mathbf \Phi$ (per  (\ref{eqn:gamma_U}) and (\ref{eqn:y_U})), and $\gamma_R$  in (\ref{eqn:obj2}) is a fourth order function of  $\mathbf \Phi$ (per  (\ref{eqn:gamma_R}) and (\ref{eqn:y_R})). 
Converting the non-convex $\gamma_R$ term into a proper form that makes $(\mathrm P_{\Phi})$  efficiently solvable is the key to  the second sub-problem.

\subsubsection{Solve for $\mathbf \Phi$ with single minorization} \label{sec:msdr}
The fourth order function of matrix $\mathbf \Phi$ cannot be optimized directly. To ensure mathematical tractability, we can at first re-write the objective function as a function of $\boldsymbol \phi
\buildrel \triangle \over = 
\text{diag} (\mathbf \Phi)$, which is a column vector  containing the diagonal elements of $\mathbf \Phi$. We also define $\mathbf R_{\phi} = \boldsymbol \phi \boldsymbol \phi^T$. Since  $\gamma_R$ is quartic in $\mathbf \Phi$, it is quartic in $\boldsymbol \phi$, and quadratic in $\mathbf R_{\phi}$. The radar SNR, $\gamma_R$, can be transformed into a linear form of $\mathbf R_{\phi}$, so that the second sub-problem $(\mathrm P_{\Phi})$ is solvable. The state-of-the-art bounding technique, MM \cite{Sun2017majorization}, is a good candidate for the transform. Minorization can be used to create a  lower bound  for $\gamma_R$ that is linear in $\mathbf R_{\phi}$, i.e. 

\vspace{-5mm}
\begin{small}
\begin{eqnarray} \label{eqn:gamma_R_t}
    \!\!\!\!\!\!\!\!\!&&\gamma_R = \tr{\mathbf R_{\phi} \mathbf M \mathbf R_{\phi}^H} \nonumber \\ \!\!\!\!\!\!\!\!\!&&\geq \gamma_R^{(t)} =\tr{\mathbf R_{\phi}^{(t)} \mathbf M \mathbf R_{\phi}^H} + \tr{\mathbf R_{\phi} \mathbf M  [\mathbf R_{\phi}^{(t)}]^H} + \text{const},
\end{eqnarray}
\end{small}
\vspace{-5mm}

\noindent where $\mathbf M$ is a matrix not relevant  to $\mathbf R_{\phi}$, $\mathbf R_{\phi}^{(t)}$ is the solution obtained for $\mathbf R_{\phi}$  in the $t$-th/previous iteration, where the current iteration index is $(t+1)$.  $\boldsymbol \phi$ is then recovered from $\mathbf R_{\phi}$, and projected onto the unit modulus complex circle to ensure feasibility. The solved  $\boldsymbol \phi$ in the current iteration is plugged in the first sub-problem in the next iteration.

It is worth noting that MM \cite{Sun2017majorization} plays the same role as the first order Taylor expansion, or the successive convex approximation (SCA), when creating the bounding or surrogate function. In a practical scenario, large-scale IRS is desired to achieve high beamforming gain \cite{Najafi2021physics}. This would result in a large size variable $\boldsymbol \phi$. The SDR method would entail high complexity in this case \cite{Jiang2021intelligent}. Low-complexity IRS parameter optimization algorithms are necessary to solve problem with high dimensional variable.

\subsubsection{Solving for $\mathbf \Phi$ with low-complexity twice minorization} \label{sec:twice_minor}
Motivated by the need for low complexity algorithm design, a twice minorization method is proposed in \cite{Li2023minorization}. By applying MM once on $\gamma_R$, which is fourth order in  $\boldsymbol \phi$, a surrogate function $\gamma_R^{(t)}$ in Eq. (\ref{eqn:gamma_R_t}) is obtained, that is quadratic in  $\boldsymbol \phi$ \cite{Jiang2021intelligent}. Since   $\gamma_U$ is also quadratic of $\boldsymbol \phi$, the surrogate objective function,  $\alpha\gamma_R^{(t)} + (1-\alpha) \gamma_U$, is  quadratic in 
$\boldsymbol \phi$. The optimization problem for $\boldsymbol \phi$, or $\mathbf \Phi$, becomes quadratic programming with UMC. To bypass the high-complexity SDR, the quadratic objective can be minorized again, to get a linear surrogate function. The following formulas are applied

\vspace{-4mm}
\begin{small}
\begin{eqnarray}  \!\!\!\!\!\!\!\!\!\!\!\!\!\!\!\!\!\!\!\!\!\!\!&&\boldsymbol \phi^T \mathbf M_1 \boldsymbol \phi \geq 2 [\boldsymbol \phi^{(t)}]^T \mathbf M_1 \boldsymbol \phi - [\boldsymbol \phi^{(t)}]^T \mathbf M_1 \boldsymbol \phi^{(t)}, \nonumber \\ \!\!\!\!\!\!\!\!\!\!\!\!\!\!\!\!\!\!\!\!\!\!&& \boldsymbol \phi^H \mathbf M_2 \boldsymbol \phi \geq [\boldsymbol \phi^{(t)}]^H \mathbf M_2 \boldsymbol \phi+\boldsymbol \phi^H \mathbf M_2 \boldsymbol \phi^{(t)}-[\boldsymbol \phi^{(t)}]^H \mathbf M_2 \boldsymbol \phi^{(t)}\!\!\!,
\end{eqnarray}
\end{small}
\vspace{-4mm}

 \noindent where $\boldsymbol \phi^{(t)}$ is the solved value of $\boldsymbol\phi$ in $t$-th/previous iteration, $\mathbf M_1$ and $\mathbf M_2$ are matrices not relevant to $\boldsymbol \phi$. By applying these two minorization functions to  the objective $\alpha\gamma_R^{(t)} + (1-\alpha) \gamma_U$, a surrogate function linear in $\boldsymbol \phi$ can be obtained  as

 \vspace{-5mm}
\begin{eqnarray}
    g(\boldsymbol \phi) = \Re \{ \boldsymbol{\phi}^H \boldsymbol \nu + \boldsymbol{\phi}^T \boldsymbol \eta \},
\end{eqnarray}
\noindent where $\boldsymbol \nu$ and $\boldsymbol \eta$ are vectors not relevant to $\boldsymbol \phi$. The IRS parameter optimization problem is thereby converted to

\vspace{-5mm}
\begin{small}
\begin{subequations} \label{eqn:sub_prob_2_new2}
	\begin{eqnarray} 
		\!\!\!\!\!\!\!\!\!  ({ {\mathrm P}}_{\boldsymbol \phi}) \qquad \max_{\boldsymbol{\phi}}&&	\!\!\!\!\!	{ g}(\boldsymbol{\phi}) \label{eqn:obj_func_new2} \\ 		
		\!\!\!\!\!\!\!\!\! \mathrm{s.t.}&& \!\!\!\!\! |{\boldsymbol\phi}_{n,1}| = 1,\;\; \forall n = 1, \cdots, N, \label{eqn:unit_modu_new2}
	\end{eqnarray}
\end{subequations}
\end{small}

\noindent where $\boldsymbol\phi_{n,1}$ denotes the $n$-th element of $\boldsymbol\phi$. The solution of $({ {\mathrm P}}_{\boldsymbol \phi})$ in the $(t+1)$-th/current iteration is 
\begin{eqnarray} \label{eqn:phi_update}
	{\boldsymbol\phi}^{(t+1)} = \exp{[j \text{arg}(\boldsymbol \nu+\boldsymbol \eta^{\ast})]}.
\end{eqnarray}
\noindent Therefore, by applying minorization twice to the original fourth order objective, a linear lower bound surrogate function is obtained. Based on that, in each iteration, a closed-form solution for updating $\boldsymbol\phi$ can be derived (see  (\ref{eqn:phi_update})). Note that this update scheme  involves only  matrix multiplication and addition; it does not require an interior point method of the CVX toolbox \cite{cvx}, whose complexity increases significantly with the variable size. In addition, with the proposed twice minorization method \cite{Li2023minorization} the Gaussian randomization necessary  for the SDR method is bypassed. 
 
The reader can refer to \cite{Li2023minorization} for more details of the derivation of the twice minorization method.  Eq. (\ref{eqn:phi_update}), for updating $\boldsymbol \phi$, is a solution of the optimization problem $({ {\mathrm P}}_{\boldsymbol \phi})$ of  (\ref{eqn:sub_prob_2_new2}). That solution update method is called power method-like iteration; the monotonicity of the objective is proved in \cite{Soltanalian2014designing}. 
 The convergence of the minorization technique has been well established \cite{Wu2018transmit}.

\subsubsection{Solve for $\mathbf \Phi$ with  minorization branch and bound}
Existing optimization algorithms for IRS-aided DFRC systems tend to terminate once a local optimum point is found. How to avoid being trapped in local optima is not clear, however. The branch and bound (BnB) framework\cite{Liu2018toward}, designed for quadratic programming problems, can select a optimal solution from iteratively acquired local optimal points.
Since the design problem of (\ref{eqn:problem2}) has a fourth order objective, directly applying BnB to $(\mathrm P_{\Phi})$, is not feasible. Recall in (\ref{eqn:gamma_R_t}) of Section \ref{sec:msdr}, we have created a lower bound for the radar SNR, $\gamma_R$, using minorization \cite{Sun2017majorization}, denoted as $\gamma_R^{(t)}$, which is quadratic in $\boldsymbol \phi$. Therefore, the surrogate objective, $\alpha\gamma_R^{(t)} + (1-\alpha) \gamma_U$, which is quadratic in $\boldsymbol \phi$, is compatible with the  BnB framework. Near-optimal solution can be obtained by applying BnB to the surrogate objective. 

The minorization-BnB method proceeds as follows. We create a feasible set pool that contains feasible sets from which we try to find a optimum. Initially, the pool contains only one feasible set,
i.e., $\{\boldsymbol \phi |\phi_n \in [0, 2\pi) \, \forall n\}$ where $\boldsymbol \phi = [e^{j \phi_1},\cdots,e^{j \phi_n},\cdots,e^{j \phi_N}]^T$ and $\phi_n$ is the phase shift of the $n$-th IRS element. 
In each iteration, we choose  the set inside which the current best solution falls, from the pool,
and equally divide it into two smaller sets. These two sets are added to the pool, and the performance metric is calculated for  each of them. The sets in the pool that have low performance will be pruned to reduce the search space. This partition-search-prune process terminates when the improvement by any further search is very small. The minorization-BnB acts as a benchmark method.

\subsubsection{Solve $\mathbf \Phi$ with  low-complexity Riemannian manifold optimization} \label{sec:rmo}

The previously mentioned methods convert the original objective, which is a fourth order function of $\mathbf \Phi$ or $\boldsymbol \phi$, into a lower order quadratic, or linear form at first, with MM technique \cite{Sun2017majorization}, to facilitate the subsequent process. However,  directly optimizing the original quartic objective is not straightforward. 

The unit modulus constraints for the elements of $\boldsymbol \phi$, defines a complex circle manifold, which is highly non-convex in the Euclidean space sense. Methods designed for the Euclidean space, use Euclidean gradient as the direction of search. This will produce solutions that are off the complex circle manifold, which does not guarantee  feasibility.

Riemannian manifold optimization (RMO)  performs gradient descent, or ascent, on the complex circle manifold. RMO can guarantee a  feasible  solution for the IRS parameter  during the optimization process.
In each iteration, the Euclidean gradient is projected to the tangent space of the current solution point. This projection, known as the Riemannian gradient, is the direction of search in the update of the solution. Afterwards, the updated solution is retracted to the unit modulus complex circle, so that the solution feasibility is ensured in the process of solution update. Let us denote the original fourth order objective function as $f(\boldsymbol \phi)=\alpha\gamma_R + (1-\alpha) \gamma_U$, and the Euclidean gradient at current point $\boldsymbol \phi_t$ as $\nabla f(\boldsymbol{\phi}_t)$. The Riemannian gradient can be computed as 

\vspace{-6mm}
\begin{eqnarray}    \mathrm{grad}_{\boldsymbol\phi_t}f = \nabla f(\boldsymbol{\phi}_t) - \mathrm{Re}\{\nabla f(\boldsymbol{\phi}_t) \circ \boldsymbol{\phi}_t^{\ast}\}\circ \boldsymbol{\phi}_t.
\end{eqnarray}
\vspace{-6mm}

Afterwards, the value of $\boldsymbol{\phi}$ is updated as
\begin{eqnarray} \label{eqn:theta_update}
\boldsymbol{\phi}_{t+1} = (\boldsymbol{\phi}_{t} + \delta_t \mathrm{grad}_{\boldsymbol\phi_t}f) \circ {|\boldsymbol{\phi}_{t} + \delta_t \mathrm{grad}_{\boldsymbol\phi_t}f|}^{-1},
\end{eqnarray}
\noindent where $\delta_t$ represents the step size determined by the Armijo rule \cite{Liu2018toward}; the part after $\circ$ represents the element-wise normalization, which retracts the updated solution to the unit modulus complex circle. This process of updating the value of $\boldsymbol{\phi}$ terminates when convergence criterion is met.

RMO can directly optimize the objective containing polynomials of high-order without creating lower-order approximations, surrogate functions, or bounds. The  original objective consists of a fourth order function of  $\boldsymbol{\phi}$, i.e., $f_4(\boldsymbol{\phi}) = h(\boldsymbol{\phi}) h^{\ast}(\boldsymbol{\phi})$, where $h(\boldsymbol{\phi})$ is a quadratic function of $\boldsymbol{\phi}$. The product rule can be applied to calculate the gradient of $f_4(\boldsymbol \phi)$
\begin{eqnarray}
    \nabla f_4(\boldsymbol{\phi}) =  \nabla [h(\boldsymbol{\phi}) h^{\ast}\!(\boldsymbol{\phi})] \!=\! h^{\ast}\!(\boldsymbol{\phi}) \nabla h(\boldsymbol{\phi}) \!+\! h(\boldsymbol{\phi}) \nabla h^{\ast}\!(\boldsymbol{\phi}).
\end{eqnarray}
We refer \cite{Li2022_sam} to readers  for more details of  applying RMO to the IRS parameter design problem, which has high order objective and is subject to unit modulus constraints for the elements of variable. In addition, the proposed RMO is also low-complexity closed-form expression based solution, like the twice minorization method \cite{Li2023minorization} in Section \ref{sec:twice_minor}, both of which scale well with the increase of the variable size. 

\subsubsection{Subsection summary}
In this subsection, we have presented  several 
methods for solving the second sub-problem, i.e., the design of  $\mathbf{\Phi}$ or $\boldsymbol \phi$, when the radar precoder $\mathbf{w}$ is fixed. These methods include  minorization-SDR \cite{Jiang2021intelligent}, twice minorization \cite{Li2023minorization}, minorization-BnB, and RMO \cite{Li2022_sam}. The minorization-SDR \cite{Jiang2021intelligent} of Section \ref{sec:msdr} is that,
the fourth order objective is first transformed into a quadratic function via minorization or MM \cite{Sun2017majorization}, which forms a quadratic programming subject to UMC. The transformed quadratic problem is solved by the classical SDR method \cite{Luo2010semidefinite}. 

One attempt to reduce complexity is to  bypass the quadratic programming, which inevitably invokes numerical solvers like CVX \cite{cvx}.
In Section \ref{sec:twice_minor}, minorization is applied to the fourth order objective to  degrade it to a quadratic one, which is degraded again by minorization to a linear one, which results in a simpler solution. 
The RMO method \cite{Li2022_sam} in Section \ref{sec:rmo} also bypasses the quadratic programming. 



\begin{figure}[!t]\centering\vspace{-0mm}
	\includegraphics[width=0.28\textwidth]{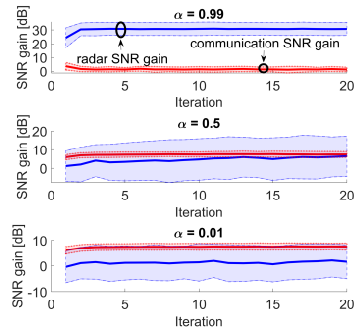}\vspace{-3 mm}
	\caption{Convergence of SNR gains for radar and communication,  $P_R = 30$\,dBm, $N_T=16$, $N=36$.}
	\label{fig:Fig_gain_converge}\vspace{-3mm}
\end{figure}

\section{NUMERICAL RESULTS}
In this section, we provide numerical results to show the trade-off between the radar and communication metrics by varying the weight factor for radar SNR ($\alpha$).  In addition, the scalability of the two presented low-complexity algorithms, i.e. twice minorization \cite{Li2023minorization} and RMO \cite{Li2022_sam}, are discussed.

Fig. \ref{fig:Fig_gain_converge} illustrates the impact of the weight factor $\alpha$ on the radar SNR. The convergence of the SNR gains for both radar and communication functions is also shown. The solid line denotes the SNR gain averaged over $50$ iterations, and the shaded area around the mean represents the variance among different realizations. The blue color is assigned to the radar metric, and the red to the communication metric. It is observed that, by decreasing $\alpha$ 
from $0.99$ to  $0.01$, the radar SNR gain  drops from around $30$\,dB to around $0$\,dB, and the communication SNR gain  increases from around $0$\,dB to around $7$\,dB. This shows that the radar SNR is more sensitive, as compared to the communication SNR, the reason being that the former is a fourth order function of  $\mathbf{\Phi}$, while the latter is a quadratic function of $\mathbf{\Phi}$.

\begin{figure}[!t]\centering\vspace{-0mm}
	\includegraphics[width=0.28\textwidth]{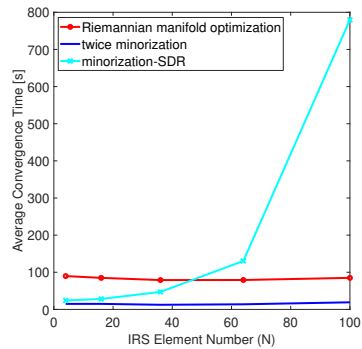}
	\caption{Average convergence time versus number of IRS elements ($N$),  $P_R = 30$\,dBm, $N_T=16$, $\alpha=0.9$.}
	\label{fig:Fig_runtime_vs_N}\vspace{-3mm}
\end{figure}

Fig. \ref{fig:Fig_runtime_vs_N} shows  the average time to converge for the different algorithms considered. It is observed that the Riemannian manifold optimization \cite{Li2022_sam} of Section \ref{sec:rmo}, and the twice minorization \cite{Li2023minorization} of  Section \ref{sec:twice_minor}, scale well with IRS size ($N$). This shows the benefit of the closed-form expression based solution update scheme, compared to numerical solvers based method, for problems with large size variable $(N)$.
  
\vspace{2mm}
\section{CONCLUSIONS}
\label{sec:conclusions}
In this paper, 
we have presented several methods of designing the IRS parameter matrix for IRS DFRC systems. The  classical method is applying the MM technique to create a solvable quadratic programming. One low-complexity design is creating a linear surrogate objective function by using MM twice. Another low-complexity method is by using Riemannian manifold optimization directly on the original fourth order objective. In addition, branch and bound framework based  design has been  presented that achieves a near-optimal solution  chosen from multiple iteratively obtained local optima.

\clearpage


\balance


\vspace{12pt}

\bibliographystyle{IEEEtran}
\bibliography{IEEEabrv,References,Ref2}

\end{document}

%% file: System_model_one_user.pdf_tex
\begingroup%
  \makeatletter%
  \providecommand\color[2][]{%
    \errmessage{(Inkscape) Color is used for the text in Inkscape, but the package 'color.sty' is not loaded}%
    \renewcommand\color[2][]{}%
  }%
  \providecommand\transparent[1]{%
    \errmessage{(Inkscape) Transparency is used (non-zero) for the text in Inkscape, but the package 'transparent.sty' is not loaded}%
    \renewcommand\transparent[1]{}%
  }%
  \providecommand\rotatebox[2]{#2}%
  \newcommand*\fsize{\dimexpr\f@size pt\relax}%
  \newcommand*\lineheight[1]{\fontsize{\fsize}{#1\fsize}\selectfont}%
  \ifx\svgwidth\undefined%
    \setlength{\unitlength}{531.44648977bp}%
    \ifx\svgscale\undefined%
      \relax%
    \else%
      \setlength{\unitlength}{\unitlength * \real{\svgscale}}%
    \fi%
  \else%
    \setlength{\unitlength}{\svgwidth}%
  \fi%
  \global\let\svgwidth\undefined%
  \global\let\svgscale\undefined%
  \makeatother%
  \begin{picture}(1,0.42687839)%
    \lineheight{1}%
    \setlength\tabcolsep{0pt}%
    \put(0,0){\includegraphics[width=\unitlength,page=1]{System_model_one_user.pdf}}%
    \put(0.32803735,0.12354536){\color[rgb]{0,0,0}\makebox(0,0)[lt]{\lineheight{0}\smash{\begin{tabular}[t]{l}$\mathbf g$\end{tabular}}}}%
    \put(0.15342076,0.3048945){\color[rgb]{0,0,0}\makebox(0,0)[lt]{\lineheight{0}\smash{\begin{tabular}[t]{l}$\large{\text {IRS}}$\end{tabular}}}}%
    \put(0,0){\includegraphics[width=\unitlength,page=2]{System_model_one_user.pdf}}%
    \put(0.54698472,0.20865845){\color[rgb]{0,0,0}\makebox(0,0)[lt]{\lineheight{0}\smash{\begin{tabular}[t]{l}$\mathbf H_{dl}$\end{tabular}}}}%
    \put(0,0){\includegraphics[width=\unitlength,page=3]{System_model_one_user.pdf}}%
    \put(0.5537353,0.05808369){\color[rgb]{0,0,0}\makebox(0,0)[lt]{\lineheight{0}\smash{\begin{tabular}[t]{l}${\text {DFRC}}$\end{tabular}}}}%
    \put(0.56137101,0.00843944){\color[rgb]{0,0,0}\makebox(0,0)[lt]{\lineheight{0}\smash{\begin{tabular}[t]{l}${\text {radar}}$\end{tabular}}}}%
    \put(0.70599365,0.38456195){\color[rgb]{0,0,0}\makebox(0,0)[lt]{\lineheight{0}\smash{\begin{tabular}[t]{l}$\large{\text {target}}$\end{tabular}}}}%
    \put(0,0){\includegraphics[width=\unitlength,page=4]{System_model_one_user.pdf}}%
    \put(0.19036861,0.18181133){\color[rgb]{0,0,0}\makebox(0,0)[lt]{\lineheight{0}\smash{\begin{tabular}[t]{l}$\mathbf f$\end{tabular}}}}%
    \put(-0.00177568,0.04142523){\color[rgb]{0,0,0}\makebox(0,0)[lt]{\lineheight{0}\smash{\begin{tabular}[t]{l}${\text {user}}$\end{tabular}}}}%
    \put(0,0){\includegraphics[width=\unitlength,page=5]{System_model_one_user.pdf}}%
    \put(0.40329917,0.14909537){\color[rgb]{0,0,0}\makebox(0,0)[lt]{\lineheight{0}\smash{\begin{tabular}[t]{l}$\mathbf H_{ul}$\end{tabular}}}}%
  \end{picture}%
\endgroup%

%% file: IRS_DFRC_tutorial.bbl
\begin{thebibliography}{10}
\providecommand{\url}[1]{#1}
\csname url@samestyle\endcsname
\providecommand{\newblock}{\relax}
\providecommand{\bibinfo}[2]{#2}
\providecommand{\BIBentrySTDinterwordspacing}{\spaceskip=0pt\relax}
\providecommand{\BIBentryALTinterwordstretchfactor}{4}
\providecommand{\BIBentryALTinterwordspacing}{\spaceskip=\fontdimen2\font plus
\BIBentryALTinterwordstretchfactor\fontdimen3\font minus
  \fontdimen4\font\relax}
\providecommand{\BIBforeignlanguage}[2]{{%
\expandafter\ifx\csname l@#1\endcsname\relax
\typeout{** WARNING: IEEEtran.bst: No hyphenation pattern has been}%
\typeout{** loaded for the language `#1'. Using the pattern for}%
\typeout{** the default language instead.}%
\else
\language=\csname l@#1\endcsname
\fi
#2}}
\providecommand{\BIBdecl}{\relax}
\BIBdecl

\bibitem{Li2016optimum}
B.~Li, A.~P. Petropulu, and W.~Trappe, ``Optimum co-design for spectrum sharing
  between matrix completion based {MIMO} radars and a {MIMO} communication
  system,'' \emph{{IEEE} Trans. Signal Process.}, vol.~64, no.~17, pp.
  4562--4575, 2016.

\bibitem{Zhang2021anoverview}
J.~A. Zhang, F.~Liu, C.~Masouros, R.~W. Heath, Z.~Feng, L.~Zheng, and
  A.~Petropulu, ``An overview of signal processing techniques for joint
  communication and radar sensing,'' \emph{IEEE Journal of Selected Topics in
  Signal Processing}, vol.~15, no.~6, pp. 1295--1315, 2021.

\bibitem{Liu2020joint}
F.~Liu, C.~Masouros, A.~P. Petropulu, H.~Griffiths, and L.~Hanzo, ``Joint radar
  and communication design: Applications, state-of-the-art, and the road
  ahead,'' \emph{{IEEE} Trans. Commun.}, vol.~68, no.~6, pp. 3834--3862, 2020.

\bibitem{Xu2023abandwidth}
Z.~Xu and A.~Petropulu, ``A bandwidth efficient dual-function radar
  communication system based on a {MIMO} radar using {OFDM} waveforms,''
  \emph{{IEEE} Trans. Signal Process.}, vol.~71, pp. 401--416, 2023.

\bibitem{Wu2019intelligent}
Q.~Wu and R.~Zhang, ``Intelligent reflecting surface enhanced wireless network
  via joint active and passive beamforming,'' \emph{{IEEE} Trans. Wireless
  Commun.}, vol.~18, no.~11, pp. 5394--5409, 2019.

\bibitem{Wu2020towards}
------, ``Towards smart and reconfigurable environment: Intelligent reflecting
  surface aided wireless network,'' \emph{{IEEE} Commun. Mag.}, vol.~58, no.~1,
  pp. 106--112, 2020.

\bibitem{Buzzi2022foundations}
S.~Buzzi, E.~Grossi, M.~Lops, and L.~Venturino, ``Foundations of {MIMO} radar
  detection aided by reconfigurable intelligent surfaces,'' \emph{{IEEE} Trans.
  Signal Process.}, vol.~70, pp. 1749--1763, 2022.

\bibitem{Li2023minorization}
Y.-K. Li and A.~Petropulu, ``Minorization-based low-complexity design for
  {IRS}-aided {ISAC} systems,'' \emph{arXiv:2302.11132}, 2023.

\bibitem{Li2022_sam}
Y.~Li and A.~Petropulu, ``Dual-function radar-communication system aided by
  intelligent reflecting surfaces,'' in \emph{2022 IEEE 12th Sensor Array and
  Multichannel Signal Processing Workshop (SAM)}, 2022, pp. 126--130.

\bibitem{Jiang2021intelligent}
Z.-M. Jiang, M.~Rihan, P.~Zhang, L.~Huang, Q.~Deng, J.~Zhang, and E.~M.
  Mohamed, ``Intelligent reflecting surface aided dual-function radar and
  communication system,'' \emph{{IEEE} Syst. J.}, pp. 1--12, 2021.

\bibitem{Liu2022joint}
R.~Liu, M.~Li, Y.~Liu, Q.~Wu, and Q.~Liu, ``Joint transmit waveform and passive
  beamforming design for {RIS}-aided {DFRC} systems,'' \emph{{IEEE} J. Sel.
  Topics Signal Process.}, vol.~16, no.~5, pp. 995--1010, 2022.

\bibitem{Sun2017majorization}
Y.~Sun, P.~Babu, and D.~P. Palomar, ``Majorization-minimization algorithms in
  signal processing, communications, and machine learning,'' \emph{{IEEE}
  Trans. Signal Process.}, vol.~65, no.~3, pp. 794--816, 2017.

\bibitem{Najafi2021physics}
M.~Najafi, V.~Jamali, R.~Schober, and H.~V. Poor, ``Physics-based modeling and
  scalable optimization of large intelligent reflecting surfaces,''
  \emph{{IEEE} Trans. Commun.}, vol.~69, no.~4, pp. 2673--2691, 2021.

\bibitem{Li2017_coexistence}
B.~Li and A.~P. Petropulu, ``Joint transmit designs for coexistence of {MIMO}
  wireless communications and sparse sensing radars in clutter,'' \emph{{IEEE}
  Trans. Aerosp. Electron. Syst.}, vol.~53, no.~6, pp. 2846--2864, 2017.

\bibitem{Luo2010semidefinite}
Z.-q. Luo, W.-k. Ma, A.~M.-c. So, Y.~Ye, and S.~Zhang, ``Semidefinite
  relaxation of quadratic optimization problems,'' \emph{{IEEE} Signal Process.
  Mag.}, vol.~27, no.~3, pp. 20--34, 2010.

\bibitem{cvx}
M.~Grant and S.~Boyd, ``{CVX}: Matlab software for disciplined convex
  programming, version 2.1,'' \url{http://cvxr.com/cvx}, Mar. 2014.

\bibitem{Soltanalian2014designing}
M.~Soltanalian and P.~Stoica, ``Designing unimodular codes via quadratic
  optimization,'' \emph{{IEEE} Trans. Signal Process.}, vol.~62, no.~5, pp.
  1221--1234, 2014.

\bibitem{Wu2018transmit}
L.~Wu, P.~Babu, and D.~P. Palomar, ``Transmit waveform/receive filter design
  for {MIMO} radar with multiple waveform constraints,'' \emph{{IEEE} Trans.
  Signal Process.}, vol.~66, no.~6, pp. 1526--1540, 2018.

\bibitem{Liu2018toward}
F.~Liu, L.~Zhou, C.~Masouros, A.~Li, W.~Luo, and A.~Petropulu, ``Toward
  dual-functional radar-communication systems: Optimal waveform design,''
  \emph{{IEEE} Trans. Signal Process.}, vol.~66, no.~16, pp. 4264--4279, 2018.

\end{thebibliography}
